\documentclass[showpacs,aps,prb,twocolumn]{revtex4-1}  

\usepackage{amsmath,amssymb}
\usepackage{graphicx}
\PassOptionsToPackage{caption=false}{subfig} 
\usepackage{subfig}
\usepackage{verbatim}
\usepackage{soul} 
\usepackage{color}
\usepackage{hyperref}

\usepackage{dsfont}

\usepackage{ulem}

\usepackage{grffile}

\def\mbs#1{\mbox{\scriptsize #1}}

\def\nnn{\nonumber \\}
\def\nn{\nonumber}

\def\d{\operatorname{d}}

\def\op#1{\mathinner{\hat{#1}}}

\newcommand\dop[2]{\op{#1}_{#2}^{\dagger}}

\def\expval#1{\mathinner{\langle{#1}\rangle}}

\DeclareMathOperator{\e}{e}

\newcommand{\comm}[2]{\left[#1,#2\right]}
\newcommand{\anticomm}[2]{\left\{#1,#2\right\}}

\def\spup{\uparrow}
\def\spdo{\downarrow}

\def\ket#1{\mathinner{|{#1}\rangle}}

\renewcommand{\imath}{\mathbf{i}}

\def\customvec#1{\mbox{\boldmath$#1$}}
\renewcommand{\vec}{\customvec}

\def\mat#1{\mathbf{#1}}

\def\transpose#1{#1^\mathrm{T}}

\newcommand{\nsigma}{\bar{\sigma}}

\begin{document}

\title{Transport and semiclassical dynamics of coupled quantum dots interacting with a local magnetic moment}

\author{Klemens Mosshammer}
\email{klemens@itp.tu-berlin.de}
\author{Gerold Kiesslich}
\email{gerold.kiesslich@tu-berlin.de}
\author{Tobias Brandes}
\affiliation{Institut f\"ur Theoretische Physik, Technische Universit\"at Berlin, Hardenbergstr. 36, D-10623 Berlin, Germany}


\begin{abstract}
We present a theory of magnetotransport through a system of two coupled electronic orbitals,
where the electron spin interacts with a (large) local magnetic moment via an exchange interaction. 
For the physical realization of such a setup we have in mind, for example, semiconductor quantum dots coupled to an ensemble
of nuclear spins in the host material or molecular orbitals coupled to a local magnetic moment. 
Using a semiclassical approximation, we derive a set
of Ehrenfest equations of motion for the electron density matrix and the mean value of the external
spin (Landau equations): Due to the spin coupling they turn out to be nonlinear and, importantly,
also coherences between electron states with different spin directions need to be considered. 
The electronic spin-polarized leads are implemented in form of a Lindblad-type dissipator in the infinite
bias limit. 
We have solved this involved dynamical system numerically for various isotropic and
anisotropic coupling schemes.
For isotropic spin coupling and spin-polarized  leads we study the effect of  current-induced magnetization of the attached spin
and compare this with a single quantum dot setup.
We further demonstrate that an anisotropic coupling can lead to a rich
variety of parametric oscillations in the average current reflecting the complicated
interplay between the Larmor precession of the external spin and the dissipative coherent dynamics
of the electron spin.
\end{abstract}

\pacs{
73.23.Hk,  
73.63.Kv,  
75.76.+j,  
85.75.-d  
}

\maketitle
\renewcommand{\theequation}{\arabic{equation}}
\renewcommand{\thefigure}{\arabic{figure}}
\renewcommand{\thetable}{\arabic{table}}


\section{Introduction}

Electronic quantum coherence in a few-orbital conductor [such as coupled semiconductor or molecular quantum dots (QDs)]
which is subject to single-electron transfer \cite{NAZ09}, can cause intriguing
measurable effects such as 
transient current oscillations,\cite{HAY03} complete current suppression,\cite{NIL10} and enhanced current fluctuations.\cite{BAR06, Kiesslich2007}
Such types of effects have been studied theoretically and
experimentally in various  QD
geometries: in serially coupled dots,\cite{WIE03,FUJ06} 
in parallel QDs,\cite{URB09, *SCH09, *SCH09c, *KAR11} in triangular
setups,\cite{MIC06} and in coupled orbitals in single molecules.\cite{DAR09}

Of particular interest is the role of  interactions of transferred electrons 
with additional bosonic or fermionic degrees of freedom.
In serial semiconductor QDs the coupling to a phononic bath can give
rise to spontaneous emission and absorption
of phonons,\cite{FUJ98, BRA99} electronic decoherence,\cite{AGU04,Kiesslich2007} or phonon replica in the transport characteristics \cite{GNO06}.
Another type of coupling is provided by the hyperfine interaction with
nuclear spins of the surrounding material.\cite{ERL01,COI04,ERL04}
This interaction enables spin-flip transfer between the electronic and the nuclear
spin system, which resolves the  current in the spin-blockade regime of serially coupled QDs.
Moreover, in this regime it can cause coherent current oscillations with a
period of the order of seconds,\cite{ONO04} which
is believed due to dynamical polarization of the nuclear spins
addressed theoretically in Ref.~\cite{INO04, *ERL05, *INA07, *INA07a, *RUD07, *RUD10}.

In transport through molecular QDs\cite{GAL08} the coupling to two
types of degrees of freedom are of particular relevance:
molecule vibrations\cite{ZHU02,
  *KOC05, *PAA05,*HUE07, *AVR09, *HAU09, *SCH09b, *HUS10, *MET11,
  *KUM12, *TraversoZiani2011} and local magnetic moments in single molecular magnets.\cite{HEE06,JO06}
The latter establishes the field of molecular
spintronics \cite{BOG08,FER08,ZUT04} and
has attracted a large amount of theoretical work,\cite{TIM06,ROS07,TIM08,KIE09,CON10,SOT10,BAU11,Bode2012,Lopez-Monis2012,SOT12}
which has focused on a single orbital as current-carrying state.
For example, Bode {\it et al.} \cite{Bode2012} have derived Landau-Lifshitz-Gilbert-type
equations of motion for the molecular spin based on a 
nonequilibrium Born-Oppenheimer-approximation, which assumes that the
time scale of the dynamics of the local magnetic moment is much larger
than the dwell time of the electrons.
In Ref.~\cite{Lopez-Monis2012} we have treated the dynamics of the average electron spin semiclassically
in the infinite bias limit. 
In the present work, we will extend our method in Ref.~\cite{Lopez-Monis2012} for the description of an electronic setup with two
serially coupled QDs, 
which interact with the same magnetic moment.
We will show that in contrast to the single-QD case, the derived equations of motion for the electronic part will possess a much more involved structure due to
the inclusion of coherences.
The resulting coupling of the semiclassical dynamics of the attached magnetic moment with the spin dynamics of the non-equilibrium electrons
gives rise to complex transient dynamics, such as chaos, and peculiar steady-state dependencies on the initial conditions.
Furthermore, we find the phenomenon of parametric resonance in the current oscillations.
We compare the current-induced switching behavior of the attached magnetic moment in our coupled setup with the single-QD setup in Refs.~\cite{Bode2012,Lopez-Monis2012}.

The remainder of the paper is organized as follows: Sec.~\ref{sec:model} contains the model with the Hamiltonian (\ref{sec:hamilt}), 
with a discussion of the level of description in Sec.~\ref{sec:separation},
with the semiclassical approximation and derivation of the 
Ehrenfest equations of motion (\ref{sec:semiclass}), and with the final equations of motion containing the transport dissipator in Sec.~\ref{sec:transport}.  
In Sec.~\ref{sec:results} the results will be discussed: steady-state currents (\ref{sec:steady}), isotropic coupling and current-induced switching of the large spin 
(\ref{sec:isotrop}), and anisotropic coupling (\ref{sec:anisotrop}). 
Finally, we will conclude in Sec.~\ref{sec:conclude}.



\section{\label{sec:model}Model}

\subsection{\label{sec:hamilt}Hamiltonian}

 \begin{figure}[htb]
  \centering
 \includegraphics{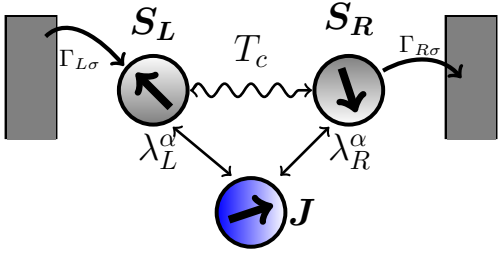}
  \caption{(Color online) Setup of two QDs mutually tunnel-coupled with strength $T_c$.
The electron spins $\vec{\op{S}}_i$ interact with a common large spin $\vec{\op{J}}$ via $\lambda_i^\alpha$. 
The QDs are attached to ferromagnetic electronic reservoirs with spin-dependent tunneling rates $\Gamma_{i\sigma}$.
}
  \label{fig:setup2} 
 \end{figure}


We consider a system of two serially coupled QDs with one orbital level each (see Fig.~\ref{fig:setup2}). 
The system is subject to an external magnetic field $B$ in the $z-$direction, which splits the QD spin levels. 
Moreover, the system is coupled to electronic leads, as well as to a large spin $\vec{\op{J}}$ with length $J$ given by 
\mbox{$\vec{\op{J}}^2\ket{m,J}=J(J+1)\ket{m,J}$}. 
The total system Hamiltonian reads as follows:

\begin{align}
\label{eq:hamiltonian}
\mathcal{H} &= \mathcal{H}_{\mbs{DQD}} + \mathcal{H}_{\mbs{J}} + \mathcal{H}_{\mbs{leads}} + \mathcal{H}_{\mbs{int}} \\
\mathcal{H}_{\mbs{DQD}} &= \sum_{\substack{i=L,R\\ \sigma=\uparrow ,\downarrow}} \varepsilon_{i} \dop{d}{i \sigma} \op{d}_{i \sigma} + T_c \sum_{i \neq j,\sigma} \dop{d}{i\sigma} \op{d}_{j\sigma} + B \sum_{i} \op{S}_i^z \nnn
 \mathcal{H}_{\mbs{J}} &= B \op{J}^z \,,\quad \quad \mathcal{H}_{\mbs{int}} = \sum_{\substack{\alpha =x,y,z\\i}} \lambda_i^\alpha \op{S}_i^\alpha \op{J}^\alpha \nnn
 \mathcal{H}_{\mbs{leads}} &=
 \sum_{l,p,\sigma}\varepsilon_{l p \sigma} \dop{c}{l p
   \sigma} \op{c}_{l p\sigma} + \sum_{lp}
 t_{lp}\sum_{\sigma} \dop{c}{l p\sigma} \op{d}_{l\sigma} + \mbox{h.c.} \,.\nn
\end{align}
Here $\mathcal{H}_{\mbs{DQD}}$ describes the double-QD system (DQD), with the tunnel coupling $T_c$ between the left and the right dot, and the energies of the orbital levels $\varepsilon_L,\varepsilon_R$, respectively. Note that our definition of $B$ comprises the Bohr magneton and $g$ factors.
For the sake of simplicity we assume identical $g$ factors for electron and large spin throughout the following.
The Coulomb repulsion between excess electrons within the QD system is assumed to be much larger than all other system and bath energies to constrain their maximal number to one (i.e., the system is operated in the Coulomb blockade regime). 

The operators $\dop{d}{i\sigma}$($\op{d}_{i\sigma}$) describe the creation(annihilation) of an electron with spin $\sigma = {\spup,\spdo}$ on the $i$th dot ($i,j,l={L,R}$); the occupation operator is defined by $\op{n}_{i \sigma} = \dop{d}{i\sigma} \op{d}_{i\sigma}$.
The relationship between these electronic operators and the $\alpha$th component of the spin operator $ \vec{\op{S}}_i$ is given by

\begin{align}
 \op{S}_i^x &= \frac{1}{2} \Big(\dop{d}{i\spup} \op{d}_{i\spdo} + \dop{d}{i\spdo} \op{d}_{i\spup} \Big) = \frac{1}{2} \Big(\op{S}_i^+  + \op{S}_i^-\Big)\nnn
 \op{S}_i^y &= \frac{1}{2 \imath} \Big(\dop{d}{i\spup} \op{d}_{i\spdo} - \dop{d}{i\spdo} \op{d}_{i\spup} \Big) = \frac{1}{2 \imath } \Big(\op{S}_i^+ - \op{S}_i^-\Big) \\
 \op{S}_i^z &= \frac{1}{2} \Big(\op{n}_{i\spup} -  \op{n}_{i\spdo} \Big)\,. \nn
\end{align}

with the usual commutation relations \mbox{$\comm{\op{S}_i^+}{\op{S}_j^-} = \delta_{i j} 2\op{S}_j^z$} and 
$\comm{\op{S}_i^z}{\op{S}_j^\pm}=\pm\delta_{i j}\op{S}_j^\pm$.

In contrast to most of other related studies, here we also allow for anisotropic coupling between electronic and large spin: $\lambda_i^\alpha\neq\lambda_i^\beta$ for $\alpha\neq\beta$.

The lead electrons are assumed to be noninteracting;  the corresponding operator $\dop{c}{lp\sigma}(\op{c}_{lp\sigma})$ creates (annihilates) an electron of momentum $p$ and spin $\sigma$ in the $l$th lead, while $t_{lp}$ is the spin-independent coupling strength between $l$th lead and $l$th dot.
%


\subsection{\label{sec:separation}Level of description and time scale separation}

The microscopic dynamics of the system described by (\ref{eq:hamiltonian}) is involved and contains a lot of information, which we are not interested in.
Therefore we will choose a level of description, where only the dynamics of single-particle observables such as components of the average electron and large spin and the 
reduced density matrix of electrons dwelling in the dots will be considered.

A rigorous derivation of the corresponding equations of motion in a controlled manner can be performed, e.g., by projective methods \cite{Rau1996} or by Keldysh-Green functions.\cite{Metelmann2012}
The price one has to pay for projecting out some ``irrelevant'' (microscopic) information is the occurence of non-Markovian terms and residual forces in the equations of motion.
One typical way to retain Markovian (time local) dynamics may be the separation of time scales present in the considered problem.

In this work we avoid this derivation path and gain the dynamics of the relevant observables by a simple semiclassical approximation instead 
(see Sec.~\ref{sec:semiclass}).
We argue phenomenologically that the time scale of the large spin precession is much slower than the electron dwell time; i.e., electron spin fluctuations do not affect the large spin dynamics and
{\it vice versa}.
Those electron spin fluctuations arise from the tunnel coupling to the electron reservoirs, which we are going to treat in the standard Born-Markov approximation.

In our Ref.~\cite{Lopez-Monis2012} we have used a technique based on Laplace transform (see Appendix~A of that reference) to combine the semiclassical Ehrenfest equations of motion for the
spin observables with the reduced density matrix for the electrons.
The same final equations of motions can be simply obtained by adding the terms based on the Lindblad master equation for dot-lead 
coupling to the spin dynamics as we have proven for the single-QD setup in Ref.~\cite{Lopez-Monis2012}.
Here, we take advantage of this observation and employ this simplified method to set up our final equations of motion (see Sec.~\ref{sec:transport}).


\subsection{\label{sec:semiclass}Semiclassical approximation and Ehrenfest equations of motion}

In order to investigate the dynamics of the coupled spin system we employ an equations of motion (EOM) technique for the expectation values of the involved spin operators. 
In general, the EOM for the expectation value of an arbitrary operator $\op{O}$ reads ($\hbar=1$)
\begin{equation}
\label{eq:eom_expval}
\frac{\d}{\d t} \expval{\op{O}} = \frac{1}{\imath}\expval{\comm{\op{O}}{\mathcal{H}}} + \expval{\frac{\partial \op{O}}{\partial t}}\,.
\end{equation}

Due to the interaction between electrons and large spin in (\ref{eq:hamiltonian}) an infinite series of coupled EOMs of higher-order spin correlators will be obtained.
In order to truncate this hierarchy we implement a semiclassical approximation on the level of the interaction Hamiltonian: 
We substitute $\op{S}_i^\alpha = \expval{\op{S}_i^\alpha}+ \delta \op{J}^\alpha$ and $\op{J}^\alpha = \expval{\op{J}^\alpha}+ \delta \op{J}^\alpha$
into the interaction Hamiltonian $\mathcal{H}_{\mbs{int}}$ and obtain
\begin{equation}
\label{eq:mean_field_interaction}
 \mathcal{H}_{\mbs{int}}^{\mbs{MF}} = \sum_{\alpha,i} \lambda_i^\alpha \left(\op{S}_i^\alpha \expval{\op{J}^\alpha} 
+ \expval{\op{S}_i^\alpha} \op{J}^\alpha - \expval{\op{S}_i^\alpha} \expval{\op{J}^\alpha} \right)\,,
\end{equation}
where the product of the spin fluctuators $\delta \op{S}_i^\alpha \delta \op{J}^\alpha$ is neglected, which is justified for 
$J \gg 1$ and $\lambda_i^\alpha /B\ll$ 1 ($\forall \alpha ,i$).
Thus, we essentially treat $\vec{\op{J}}$ as a classical object. 
Note, that since the Hamiltonian \eqref{eq:hamiltonian} does not contain any interaction of the large spin with the lead electrons or with an additional bath its 
length $J$ will be conserved on the microscopic level ($\comm{\vec{\op{J}}^2}{\mathcal{H}} = 0$).
Nevertheless, later on we will introduce a way of damping in the large spin EOMs, which is not microscopically motivated.

Using the semiclassical interaction Hamiltonian (\ref{eq:mean_field_interaction}) yields the following correlators for the electronic part of EOM  (for the sake of clarity we omit the time dependencies)

\begin{widetext}
\begin{align}
\label{eq:eom_dot_dot_correlator}
 \frac{\d}{\d t} \expval{\dop{d}{i\sigma} \op{d}_{j \sigma'}} &= \frac{\imath}{2} \Big[\left(\delta_{\sigma \spup} - \delta_{\sigma \spdo}\right) \left(B+\lambda_i^z \expval{\op{J}^z}\right) + \left(\delta_{\sigma' \spdo} - \delta_{\sigma' \spup} \right) \left(B +\lambda_j^z \expval{\op{J}^z}\right) + 2 \left(\varepsilon_i - \varepsilon_j \right) \Big] \expval{\dop{d}{i\sigma} \op{d}_{j \sigma'}}  \nnn
& + \imath T_c \left(\expval{\dop{d}{\bar{i} \sigma} \op{d}_{j \sigma'}} - \expval{\dop{d}{i\sigma} \op{d}_{\bar{j} \sigma'}} \right) \nnn
&+ \frac{\imath}{2} \left[\lambda_i^x \expval{\op{J}^x}  - \imath \left(\delta_{\sigma \spup} - \delta_{\sigma \spdo}\right) \lambda_i^y  \expval{\op{J}^y} \right] \expval{\dop{d}{i \nsigma} \op{d}_{j \sigma'}} - \frac{\imath}{2} \left[\lambda_j^x \expval{\op{J}^x} - \imath \left(\delta_{\sigma' \spdo} - \delta_{\sigma' \spup} \right) \lambda_j^y  \expval{\op{J}^y} \right] \expval{\dop{d}{i \sigma} \op{d}_{j \nsigma'}} \nnn
&-\imath \sum_p \Big[\gamma_{L p \sigma'}^* \delta_{Lj} \expval{\op{d}_{i\sigma}^{\dagger } \op{c}_{Lp\sigma'}^{}} + \gamma_{R p \sigma'}^* \delta_{Rj} \expval{\op{d}_{i\sigma}^{\dagger } \op{c}_{Rp\sigma'}^{}} - \gamma_{L p \sigma} \delta_{Li} \expval{\op{c}_{Lp\sigma}^{\dagger } \op{d}_{j\sigma'}^{}} -\gamma_{R p \sigma} \delta_{Ri} \expval{\op{c}_{Rp\sigma}^{\dagger } \op{d}_{j\sigma'}^{}}  \Big]\,,
\end{align} 
\end{widetext}
where $\bar{i}= L(R)$ for $i=R(L)$, $\nsigma = \spdo (\spup)$ for $\sigma=\spup (\spdo)$.

The EOM for the expectation values of the large spin's degree of freedom take the form of a Bloch equation completed by the electronic back-action and the phenomenological damping: 
\begin{align}
 \label{eq:eom_largespin}
 \frac{\d}{\d t} \expval{\vec{\op{J}}}(t) = \left[B\vec{e}_z + \sum_i \expval{\vec{\op{S}}_{i}'}(t) \right]  \times  \expval{\vec{\op{J}}}(t) - \gamma_J \expval{\vec{\op{J}}}(t),\nnn
\end{align}
with $\expval{\vec{\op{S}}_i'}(t) \equiv \sum_\alpha\lambda_i^\alpha \expval{\op{S}_i^{\alpha}}(t)\vec{e}_\alpha$.

So far we have derived the Ehrenfest EOM for the closed coupled spin system.
In order to pursue a microscopic derivation of spin-dependent rates for the transport between the electronic contacts and the system 
we could follow the method proposed in Ref.~\cite{Lopez-Monis2012}. 
This technique requires the direct computation of all dot-lead correlators appearing in Eq.~(\ref{eq:eom_dot_dot_correlator}):
$\expval{\op{d}_{i \sigma}^{\dagger} \op{c}_{j p \sigma'}^{}}$ and $\expval{\op{c}_{ip\sigma'}^{\dagger } \op{d}_{j \sigma}^{}}$, respectively.
However, the number of relevant dynamical variables in the DQD $\expval{\dop{d}{i\sigma} \op{d}_{j \sigma'}}(t)$ is far too large for the analytical derivation
along the lines of Appendix~A in Ref.~\cite{Lopez-Monis2012}.
Instead we make use of the fact that for the single-QD the same equations of motion can be obtained alternatively by replacing the last term of the right-hand side of
(\ref{eq:eom_dot_dot_correlator})
with terms originating from the Lindblad-type master equation. 
Even though there is no proof of equivalence we believe that those aproaches lead to the same results in the case of DQD. 
The corresponding derivation will be provided in the following section.


\subsection{\label{sec:transport}Transport master equation}

Here, we will combine the Ehrenfest EOM (\ref{eq:eom_dot_dot_correlator}) with the quantum master equation of the DQD that is not coupled to the large spin. Specifically, we use the Lindblad-type master equation for the system density matrix $\hat\rho$, which is derived by means of the standard Born-Markov-approximation \cite{Breuer2002} in the infinite bias limit:
\begin{align}
\frac{\d}{\d t}\op{\rho}(t) &= -\imath \comm{\mathcal{H}_{\mbs{DQD}}}{\op{\rho}(t)}\nnn
&-\frac{1}{2}  \sum_{\sigma} \Bigg[ \Gamma_{L\sigma}
\bigg(\anticomm{\op{d}_{L\sigma} \dop{d}{L\sigma}}{\op{\rho}(t)}\nnn
& - 2 \dop{d}{L\sigma} \op{\rho}(t) \op{d}_{L\sigma} \bigg)\nnn
&+ \Gamma_{R\sigma} \left(\anticomm{\dop{d}{R\sigma} \op{d}_{R\sigma}}{\hat\rho(t)} - 2 \op{d}_{R\sigma} \op{\rho}(t) \dop{d}{R\sigma} \right) \Bigg].\nnn
\label{eq:lindblad}
\end{align}
The matrix elements of $\op{\rho}$ are $\expval{\dop{d}{i\sigma} \op{d}_{j \sigma'}}$.
The tunnel rates are given by $\Gamma_{l\sigma}=2\pi\vert t_{l}\vert^2\rho_{l\sigma}$ with the spin-dependent density of states $\rho_{l\sigma}$ in the $l$th lead, which is approximated to be energy-independent.
The asymmetry in the density of states will be parametrized by the degree of spin polarization (see also \cite{Braun2004})
\begin{align}
 p_l\equiv \frac{\rho_{l\uparrow}-\rho_{l\downarrow}}{\rho_{l\uparrow}+\rho_{l\downarrow}}\,,
\end{align}
with $p_l\in [-1,1]$; $p_l=0$ corresponds to a nonmagnetic lead and $p_l=\pm$ 1 describes a spin up/down-polarized half-metallic ferromagnetic contact, respectively.
The tunneling rates then read $\Gamma_{l\uparrow}=\frac{1}{2}\Gamma_l(1+p_l)$,
$\Gamma_{l\downarrow}=\frac{1}{2}\Gamma_l(1-p_l)$, 
and $\Gamma_l=\Gamma_{l\uparrow}+\Gamma_{l\downarrow}$.

Now we define the density matrix $\hat\rho$ in vector form with the following subvectors (omitting the time dependence):
\begin{align}
 \vec{\rho}^\sigma &\equiv \transpose{\left(\expval{\op{\rho}_L^\sigma},\expval{\op{\rho}_R^\sigma},\expval{\op{\rho}_{LR}^\sigma},\expval{\op{\rho}_{RL}^\sigma} \right)} \nnn 
 & = \transpose{\left(\expval{\op{n}_{L\sigma}},\expval{\op{n}_{R\sigma}},\expval{\dop{d}{L\sigma} \op{d}_{R\sigma}},\expval{\dop{d}{R\sigma} \op{d}_{L\sigma}} \right)} \nnn
 \vec{\xi}_e &\equiv \transpose{\left(\expval{\op{S}_L^+},\expval{\op{S}_L^-},\expval{\op{S}_R^+},\expval{\op{S}_R^-}\right)} \\ 
 \vec{\xi}_u &\equiv \transpose{\left(\expval{\dop{d}{L\spup} \op{d}_{R\spdo}}, \expval{\dop{d}{R\spdo} \op{d}_{L\spup}} ,\expval{\dop{d}{R\spup} \op{d}_{L\spdo}} ,\expval{\dop{d}{L\spdo} \op{d}_{R\spup}} \right)} \nn
\end{align}

and the vacuum operator $\op{\rho}_0\equiv\mathds{1}-\hat{N}$ with the total electron number operator $\hat{N}=\sum_{l\sigma}\hat{n}_{l\sigma}$; its EOM is obtained by (\ref{eq:lindblad}) and reads
\begin{align}
\frac{\d}{\d t} \expval{\op{\rho}_0}(t) &= -\Gamma_{L} \expval{\op{\rho}_0}(t) + \Gamma_{R\spup} \expval{\op{\rho}_R^\spup}(t)
 + \Gamma_{R\spdo} \expval{\op{\rho}_R^\spdo}(t).
\label{eq:vacuum-EOM}
\end{align}
Eventually, the combination of \eqref{eq:eom_dot_dot_correlator}, \eqref{eq:lindblad}, and \eqref{eq:vacuum-EOM} results in the  EOM of the electronic part, which we provide in a compact vector form
\begin{align}
\frac{d}{dt}\hat{\rho}(t)=\mathcal{M}\bigg(\expval{\hat{J}^x},\expval{\hat{J}^y},\expval{\hat{J}^z}\bigg)\hat{\rho}(t)
\label{eq:mastereq}
\end{align}
with $\hat\rho (t)
\equiv\big(\expval{\op{\rho}_0}(t),\vec{\rho}^\spup(t),\vec{\rho}^\spdo(t),
\vec{\xi}_e(t),\vec{\xi}_u(t)\big)^T$ and
\begin{align}
\label{eq:system_compact_matrix}
\mathcal{M}\equiv
\begin{pmatrix}
-\Gamma_L & \vec{r}^\uparrow & \multicolumn{1}{c|}{\vec{r}^\downarrow} & \vec{0}_4^T & \vec{0}_4^T\\
\vec{l}^\uparrow & \mat{L}^\uparrow & \multicolumn{1}{c|}{\mat{0}_{4,4}} & \mat{A} &
\mat{B}\\
\vec{l}^\downarrow & \mat{0}_{4,4} & \multicolumn{1}{c|}{\mat{L}^\downarrow} & -\mat{A} &
\mat{C}\\ \cline{1-3}
\vec{0}_4 & -\mat{A}^\dagger & \mat{A}^\dagger & \mat{D} & \mat{E}\\
\vec{0}_4 & -\mat{B}^\dagger & -\mat{C}^\dagger & \mat{E} & \mat{F}\\
\end{pmatrix}
\end{align}
where $\mat{0}_{4,4}$ is the (4,4)-zero-matrix, $\vec{0}_4$ is the 4-dimensional zero-vector, $\vec{l}^\sigma\equiv \transpose{(\Gamma_{L\sigma},0,0,0)}$, \mbox{$\vec{r}^\sigma\equiv(0,\Gamma_{R\sigma},0,0)^T$}. 

Together with the modified Bloch equations
  \eqref{eq:eom_largespin} for the large spin
  $\langle\vec{\op{J}}\rangle$ 
this forms an involved set of nonlinear
differential equations, which describe the coupled dynamics of the
non-equilibrium QD
system and the large spin.
Note that due to the internal tunnel coupling coherences
between different spin directions in different QDs ($\xi_e$,$\xi_u$)
also need to be considered in the EOM.
The matrix on the right-hand side of \eqref{eq:system_compact_matrix} provides the
coupling structure; the detailed definition of the 4$\times$4 block matrices
$\mat{L}^\sigma,\mat{A},\mat{B},\dots\mat{F}$ can be found in Appendix \ref{sec:block_matrices}.
In particular, for vanishing coupling $\lambda_i^\alpha =$ 0 the
$\op{\rho}$ part decouples completely from the $\hat\xi$-part and from the
large spin $\langle\vec{\op{J}}\rangle$.
The resulting master equation for $\hat\rho$ describes the
unidirectional transport of electrons 
through two spin channels.\cite{Gurvitz1996,Stoof1996}
Due to the strong Coulomb blockade the electron transfer in those channels is correlated and
for finite coupling $\lambda_i^\alpha$ they mutually couple
due to spin-flip transfer between electron and large spin.

Since we consider the QD system in the infinite bias limit the average time-dependent electron current through the DQD can be calculated by the product of occupations and tunnel rates of the right QD
\begin{equation}
 \expval{\op{I}}(t) = \Gamma_{R\spup} \expval{\op{n}_R^\spup}(t) + \Gamma_{R\spdo} \expval{\op{n}_R^\spdo}(t)\,.
\end{equation}


\section{\label{sec:results}Analysis and Numerical results}

\subsection{\label{sec:steady}Steady-state currents}

\begin{figure}[htb]
\label{fig:large_spin_switch_undamped}
\includegraphics[width=0.5\textwidth,keepaspectratio=true]{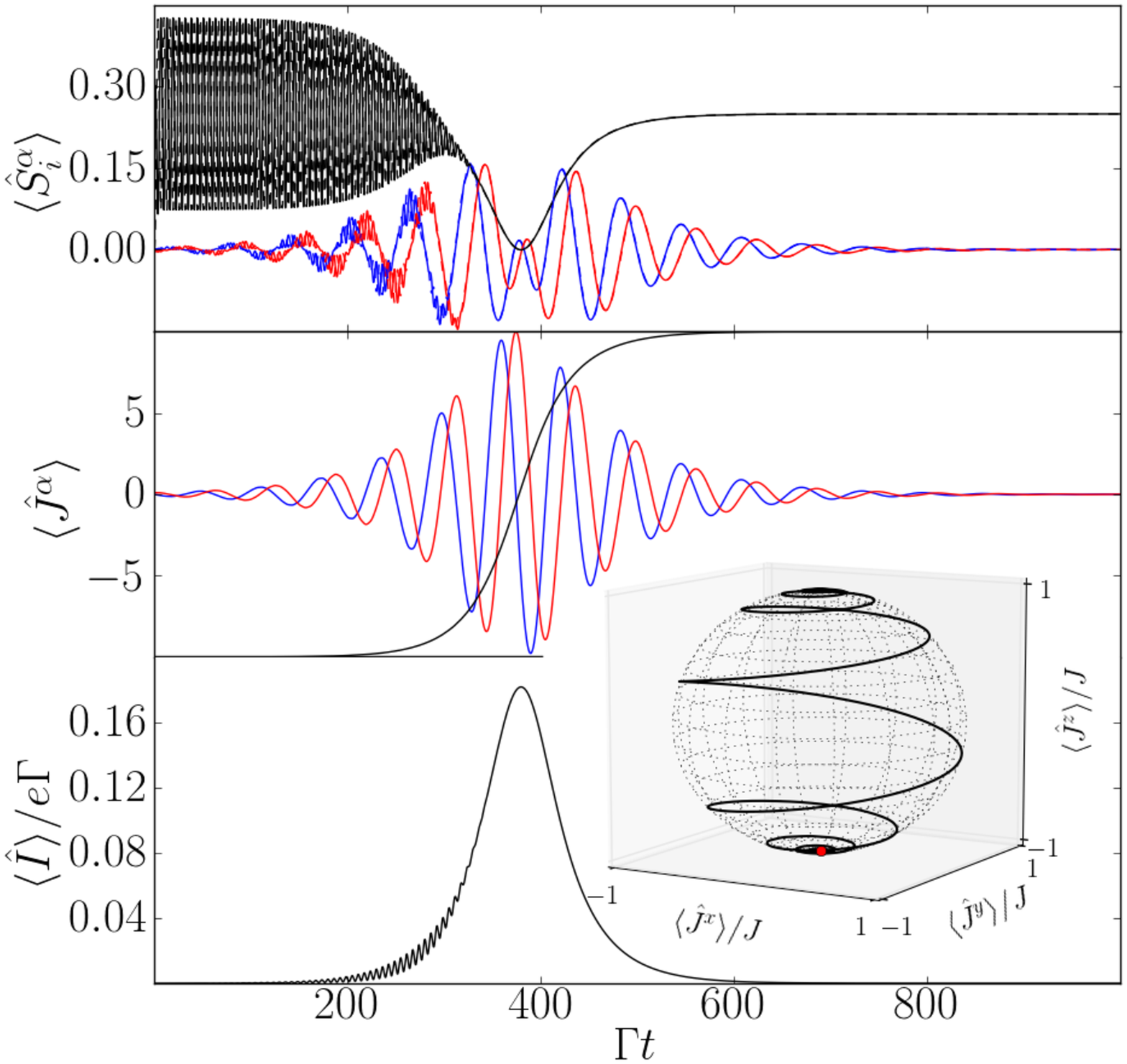}
\caption{(Color online) Current-induced magnetization of the large spin. 
(Top)  Components of the electron spin $\expval{\op{S}_{L}^x}$ (blue, solid), $\expval{\op{S}_{R}^x}$ (blue, dashed)
$\expval{\op{S}_{L}^y}$ (red, solid), $\expval{\op{S}_{L}^y}$ (red, dashed), $\expval{\op{S}_{L}^z}$ (black, solid), $\expval{\op{S}_{R}^z}$ (black, dashed). 
(Middle)  Components of the large spin: $\expval{\op{J}^x}$ (blue),  $\expval{\op{J}^y}$ (red),  $\expval{\op{J}^z}$ (black).
(Bottom)  Electronic current $\expval{I}(t)$ vs. time.
In the long-term limit one observes $\expval{\op{J}^x}=\expval{\op{J}^y}=$ 0, $\expval{\op{J}^z}=J$ and the current vanishes.
(Inset)  Evolution of the large spin as trajectory in the Bloch sphere, where the red circle depicts the initial position.
Note that the onset time depends on the initial orientation of the large spin, the more the large spin deviates from alignment with the magnetic field, 
the faster the switching process sets in.
Parameters: $p_L = -p_R = 1 $, $B/\Gamma =$ 0.1, $T_c/\Gamma=$ 0.5, $\lambda/\Gamma =$ 1, 
$\varepsilon=0$, $\gamma_J/ \Gamma=$ 0, large spin length $J=10$.}
\end{figure}

\begin{figure*}[htb]
\centering
\label{fig:pleft_pright_currents_switching_contour}
\subfloat[]{\label{fig:pleft_pright_currents_switching_contour_dqd} \includegraphics[width=0.5\textwidth,keepaspectratio=True]{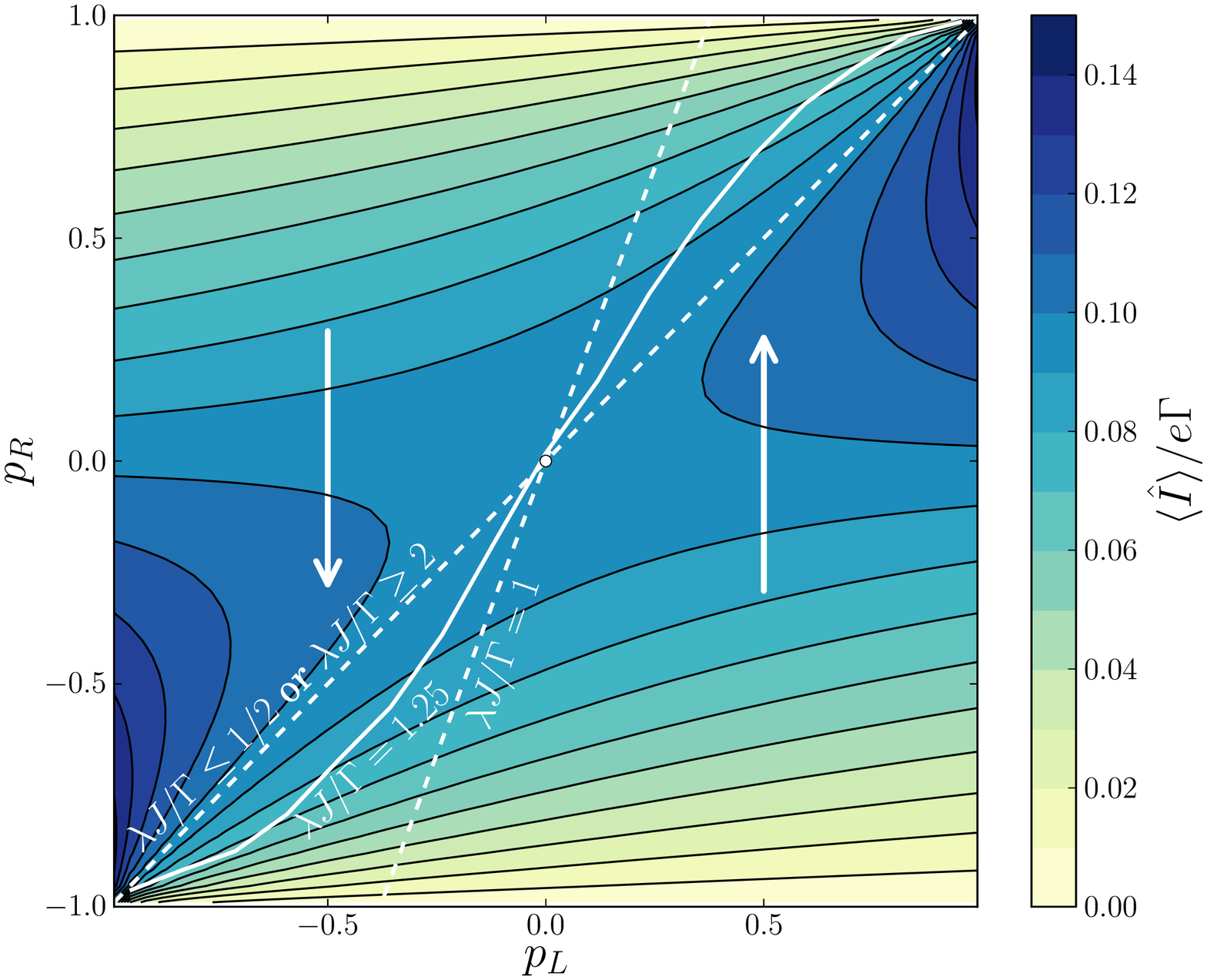}}
\subfloat[]{\label{fig:pleft_pright_currents_switching_contour_sqd} \includegraphics[width=0.5\textwidth,keepaspectratio=True]{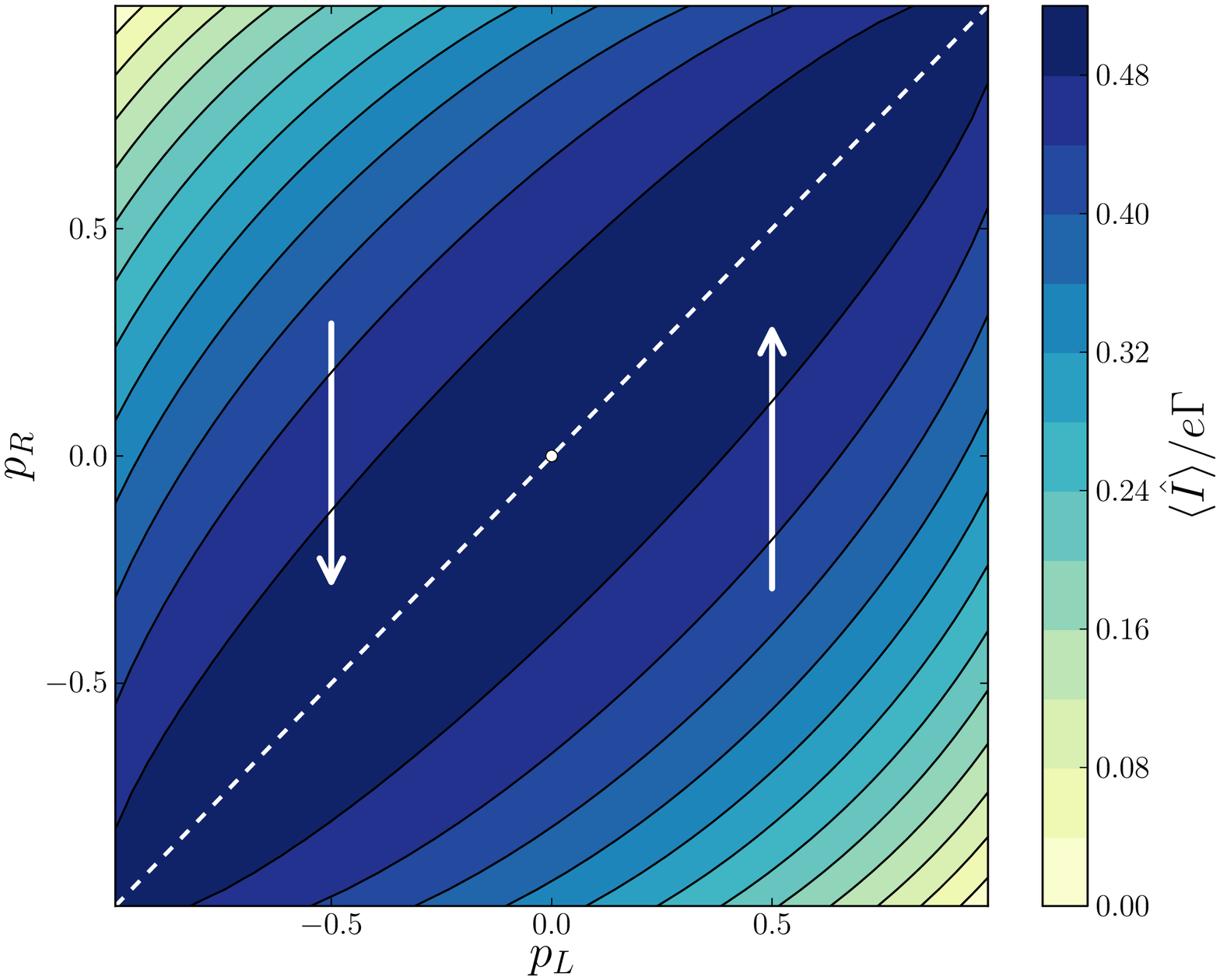}}
\caption{(Color online) Contour plots of the stationary currents vs. $p_L$ and $p_R$.
(a) Coupled QDs with parameters $\varepsilon=0$, $T_c/\Gamma =$ 0.5,  $B/\Gamma=$ 0.1.
(b) Single-QD  with the same magnetic field $B$.
The vertical arrows denote switching directions in areas divided by the straight dashed lines. 
For $p_L=p_R=0$, i.e., unpolarized contacts, and on the dashed straight lines no switching takes place. 
Note that for the DQD the transition between switching up and down depends on the product $\lambda J$: 
For $\lambda J/\Gamma \le 1/2$ and $\lambda J/\Gamma \ge$ 2 the transition takes place along the line $p_R=p_L$;
otherwise the transition line is located in the region bounded by $p_R\approx 2.6\,p_L$ for $\lambda J/\Gamma=$ 1 (dashed white line).
Note that those transitions are not linear with respect to $p_L$ as shown, e.g., for $\lambda J/\Gamma =$ 1.25 (solid white line).}
\end{figure*}

Before we will start with the discussions of the dynamics of
  large and electronic spin, we consider the steady-state currents
  in the uncoupled and coupled case.
With the interaction to the large spin turned off (i.e., $\lambda=0$)
the steady-state current can be obtained analytically and reads
%
\begin{align}
\label{eq:stationary_current}
\frac{\expval{\op{I}}}{e} = \frac{4 T_c^2 (\Gamma_{L\spdo} + \Gamma_{L\spup}) \Gamma_{R\spdo} \Gamma_{R\spup}}{
4 a\, T_c^2  + b_\spdo + b_\spup}\,,
\end{align}
where $a\equiv 2 (\Gamma_{L\spup} \Gamma_{R\spdo} + \Gamma_{L\spdo} \Gamma_{R\spup}) + \Gamma_{R\spdo}\Gamma_{R\spup}$, $b_\sigma\equiv\Gamma_{L\sigma}\Gamma_{R\bar\sigma}({\Gamma_{R\sigma}}^2 + 4 \varepsilon^2)$, and
 $\varepsilon\equiv\varepsilon_L-\varepsilon_R$ denotes the level detuning between the left and right dot levels in the system.
With the definition of the tunneling rates and the assumption of
  symmetric tunnel couplings $\Gamma_L=\Gamma_R\equiv\Gamma$ this can be
  rewritten as
\begin{align}
\frac{\expval{\op{I}}}{e\Gamma} = \frac{4T_c^2
  (1-p_R^2)}{4T_c^2[5-p_R(4p_L+p_R)]+8\varepsilon^2(1-p_Lp_R)+c\Gamma^2},
\end{align}
with $c\equiv (1+p_Lp_R)(1-p_R^2)/2$.
For symmetric polarizations $p\equiv p_L=p_R$ we further obtain 
\begin{align}
\label{eq:current_single_channel1}
\frac{\expval{\op{I}}}{e\Gamma} = \frac{4 T_c^2}{20 T_c^2 + 8 \varepsilon^2
+(1+p^2)\Gamma^2/2}\,,
\end{align}
which becomes maximal for nonmagnetic leads ($p=$ 0).
We need to note that this current expression is 
only valid for incomplete polarization ($\vert p\vert <$ 1)
since in the polarized case the steady-state version of the master equation (\ref{eq:mastereq})
is not well defined:
In particular, the relevant part of the coefficient matrix (Liouvillian superoperator) $\mathcal{M}$ (\ref{eq:system_compact_matrix})  
reduces to the left upper $N\times N$ submatrix (detached by lines) with $N=$~9.
For $|p|<$~1 its rank is $(N-1)$ and a unique steady state follows.
In contrast for $|p|=$~1 when one spin channel is completely switched off the rank reduces and more than one steady state is obtained, 
which is clearly unphysical.
The steady-state master equation is over-determined.
In order to obtain the steady state for complete
polarization we have to make use of the spin-polarized
master equation considered in Refs.~\cite{Stoof1996,Kiesslich2007} where $N=$~5.
The corresponding current reads
\begin{align}
\label{eq:current_single_channel}
\frac{\expval{\op{I}}_p}{e\Gamma} = \frac{4 T_c^2}{12 T_c^2 + 8
  \varepsilon^2 + \Gamma^2 }\,,
\end{align}
which significantly differs from
  (\ref{eq:current_single_channel1}) and leads to a  discontinuity
of the current when $\vert p\vert$ approaches one.
Therefore, the complete symmetric lead polarization provides a singular case in our description, 
which  has to be either excluded or treated with caution (see below).

\noindent
{\it Coupled spins.}-- For arbitrary spin couplings we have not been able to derive an analytical
expression for the current due to the high dimensionality of the EOM.
However, when the large spin components $\expval{\hat{J}^x}$ and $\expval{\hat{J}^y}$ approach zero in the 
long-term limit, we again obtain the steady-state current \eqref{eq:stationary_current}. 
This either occurs when the large spin is damped ($\gamma_J>$ 0) or
for isotropic spin coupling ($\lambda_i^\alpha=\lambda$, $\forall i,\alpha$) as will be discussed in
the next section.

As in the uncoupled case complete symmetric contact polarization ($\vert
p\vert =$ 1) causes pecularities, such as a dependency of the stationary current on initial conditions or multistable current behavior
(see Sec.~\ref{sec:anisotrop}).
We note, that this is also caused by the above mentioned failure of the electronic master equation and not by the semiclassical approximation
of the spin interaction.
However, the spin-polarized master equation can not be used here.

\subsection{\label{sec:isotrop}Isotropic coupling and current-induced magnetization of the large spin}
In this section we consider isotropic coupling $\lambda_i^\alpha = \lambda$ ($\forall i,\alpha$) between the electron spin and the
large spin.
We will discuss the dynamical behavior of both spins and the current for various lead polarizations $p_L$ and $p_R$.

\noindent
{\it Reverse lead polarizations.} --
For the sake of clarity we start our discussions with complete reverse polarization $p_L = - p_R = \pm 1$.
An electron transfer through the QD system only takes place when each spin-up (spin-down) electron entering the left QD is able to down-flip (up-flip) its spin state, respectively.
According to the microscopic spin coupling  $\mathcal{H}_{\mbs{int}}=\lambda\sum_j (\frac{1}{2}\op{S}_j^+\op{J}^-+\frac{1}{2} \op{S}_j^-\op{J}^++\op{S}_j^z\op{J}^z )$ this electron spin-up (spin-down) flip is accompanied with
the decrement (increment) of the large spin magnetic number by one: $m\mp$1 ($\op{J}^z\ket{m,J}=m\ket{m,J}$ with $\vert m\vert\le J$).
Once the minimal (maximal) number $\mp J$ is reached, the large spin is completely aligned with the $z-$direction  and the current vanishes.
This process is independent of the external magnetic field
and can be considered as current-induced switching \cite{Bode2012} of the attached local magnetic moment.
This effect is also known in magnetic layers described by a Landau-Lifshitz-Gilbert equation for the layer magnetization.\cite{Ralph2008}
In this approach the current-induced magnetization has been interpreted as spin-transfer torque \cite{Ralph2008,Brataas2012} acting on the magnetic moment.
Unfortunately, for our DQD we are not able to identify analogous terms in Eq.~\ref{eq:eom_largespin} due to the complex coupling structure with the electronic part.

Figure \ref{fig:large_spin_switch_undamped} shows the dynamics of the current-induced switching of $\expval{\op{\vec{J}}}$ for our DQD, 
in the case of $p_L = - p_R = \pm 1$ and no damping of the large spin $\gamma_J =$ 0.
In the transient regime, damped coherent oscillations of the electron between the QDs with frequency $2T_c$ (\mbox{$\varepsilon=$ 0}) are visible in the current and in the electron spin evolution.
For the sake of clear demonstration we have chosen the initial large spin to be reversely aligned to its final direction. 
According to our above explanation the current then is peaked around the time when the derivative of $\expval{\op{J}^z}(t)$ is maximal, in the isotropic case when $\expval{\op{J}^z} = 0$, as the spin-spin interaction and 
thus the spin-transfer becomes maximal there. 
In the long-term limit the current is given by \eqref{eq:stationary_current}.
If one assumes  a finite damping $\gamma_J>$ 0 the $z-$component of
the large spin will approach zero as well.

The inset of Fig. \ref{fig:large_spin_switch_undamped} contains the large spin evolution in the Bloch sphere -- in addition to the large spin magnetization from
$\expval{\op{J}^z} = -J$ to $+J$ a precession with frequency $B$ occurs.
The time for the magnetization reversal (switching time), however, does not depend on the external magnetic field.

Since the coupling parameter $\lambda$ provides the number of spin-flips per unit time and $\Gamma$ the number of electrons entering the left QD per unit time
the switching time is determined by their ratio: It diverges when
$\lambda /\Gamma\to$ 0 and saturates for $\lambda/\Gamma \simeq$ 1 
as long as the tunnel coupling $T_c$ between the QDs is on the order of $\Gamma$.
This provides a lower bound for the switching time, whereas for
$T_c/\Gamma\gg 1$ and  $\ll 1$ it increases.

Aside from the magnetization time scale, we observe that the onset of the switching process depends on the initial orientation of the large spin. 
In particular, if it is nearly aligned with magnetic field ($\expval{\op{J}^z}_0\approx J$) 
the spin-flip rate becomes significantly diminished, since the corresponding coupling is proportional to 
$\lambda (\expval{\op{J}^x}+\expval{\op{J}^y})$ [see \eqref{eq:mean_field_interaction}].
This leads to a slowing down of the magnetization onset; in the limit $\expval{\op{J}^z}_0=J$ the large spin keeps his initial orientation.

\noindent
{\it Arbitrary lead polarizations.} --
So far we have addressed complete reverse lead polarizations.
However, switching also takes place for arbitrary $p_L$ and $p_R$ as shown in  \ref{fig:pleft_pright_currents_switching_contour}(a).
The direction of the spin torque is schematically depicted by vertical arrows depending on $p_L$ and $p_R$ together with the steady-state current \eqref{eq:stationary_current}
as contour plot.
For $p_L< p_R$ ($p_L> p_R$) the large spin will switch to the spin-down (spin-up) direction, respectively.
Note that this holds only for parameters $\lambda J/\Gamma \le \frac{1}{2}$ or $\lambda J/\Gamma \ge 2$, respectively. 
If $\frac{1}{2} < \lambda J/\Gamma < 2$ the transition between up- and down-switching lies in the region bounded
by $p_R=p_L$  and $p_R\approx 2.6\,p_L$ for $\lambda J/\Gamma = $ 1. 
In this area the transition lines are not linear with respect to $p_L$, as shown for $\lambda J/\Gamma=$ 1.25 in Fig.~\ref{fig:pleft_pright_currents_switching_contour}(a).

\begin{figure}[t!]
  \centering
  \includegraphics[width=0.5\textwidth,keepaspectratio=True]{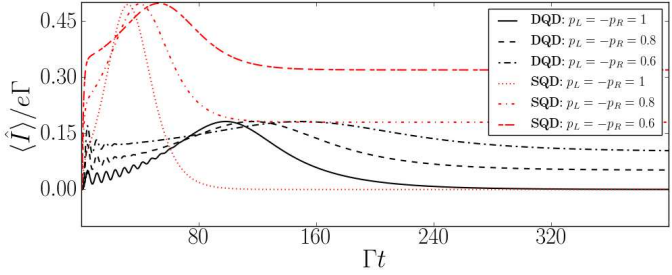}
  \caption{(Color online) Time-dependent currents in single-QD and DQD for different contact polarizations $p_L$ and $p_R$ during the process of magnetization reversal.
The currents possess a maximum when $\expval{\op{J}^z} = 0$.
With decreasing $\vert p_L\vert=\vert p_R\vert$ the time for magnetization switching increases.
In DQD the switching time is always larger than for the single-QD.
Parameters: $\lambda /\Gamma =$~1, $T_c/\Gamma =$~0.5, $B/\Gamma=$~0.1 }
  \label{fig:compare_switching_SQD_DQD_currents} 
 \end{figure}

The switching time is also affected by the choice of lead polarizations. 
Given that $|p_L| = |p_R|=|p|$, the switching time decreases with decreasing $|p|$, as depicted in Fig.\ref{fig:compare_switching_SQD_DQD_currents}. Given a fixed left lead polarization $|p_L|<1$, while $p_R$ is variable, the switching time has a lower bound for the minimum of $-p_L/p_R$. On the other hand the spin transfer is increased as the left lead polarization is increased. Thus, the switching time is minimized for complete reverse lead polarization $p_L = -p_R = \pm 1$.

In the nonmagnetic case $p_L = p_R = 0$ [origins in Figs.~\ref{fig:pleft_pright_currents_switching_contour}(a) and (b)] no switching
takes place and the electron spin state ends up in a completely mixed state. 
When both QDs are presumed to be initially unoccupied the electronic spin state is completely mixed for the entire evolution.
Consequently, the large spin evolves independently of the electron spins. 
This enables an effective description of the electron 
spin dynamics presented in Appendix \ref{sec:no_backaction}.
The situation, however, changes when the QDs are initially occupied. 
During the decay of the electronic state into a complete mixture the large spin evolution becomes affected and 
possesses transient oscillations before it runs into the back-action free precession for \mbox{$\gamma_J=$ 0}.
The frequency of the transient oscillations depend on the parameter $\lambda/\Gamma$ while their duration is governed by $\Gamma$.           
It follows that in the undamped case the large spin steady state depends on the initial QD occupation.
{\it Isotropic coupling with difference between left and right site} --
For a coupling scheme where the components of one electronic spin are isotropic, i.e., $\lambda_i^\alpha = \lambda_i$ ($\forall\alpha$)
but the coupling differs between the two electronic sites, i.e, $\lambda_L \ne \lambda_R$,
we observe in principle the same current-induced magnetization behavior as before.
However, the difference $\Delta \lambda\equiv \lambda_L - \lambda_R$ has an impact on the specific behavior, for instance, on the switching times.
The more the couplings differ, i.e., for increasing $|\Delta\lambda|$, the longer the switching takes. 
Given that the coupling strengths are on the order of magnitude of the tunneling rates, the sign of $\Delta \lambda$ is of subordinate importance.

\subsection{\label{sec:comparison_sqd}Comparison with single-QD}
In the following we will compare the phenomenon of current-induced switching in a single-QD and in the DQD in the same semiclassical description.
The total Hamiltonian for the single-QD (SQD) setup reads
\begin{align}
\label{eq:sqd_hamiltonian}
\mathcal{H} &= \mathcal{H}_{\mbs{SQD}} + \mathcal{H}_{\mbs{J}}  +
\mathcal{H}_{\mbs{int}} +\mathcal{H}_{\mbs{leads}}  \\
\mathcal{H}_{\mbs{SQD}} &= \sum_{\sigma=\uparrow ,\downarrow} \varepsilon \dop{d}{\sigma} \op{d}_{\sigma} + B \op{S}^z \nnn
 \mathcal{H}_{\mbs{leads}} &=
 \sum_{l,p,\sigma}\varepsilon_{l p \sigma} \dop{c}{l p
   \sigma} \op{c}_{l p\sigma} + \sum_{lp}
 t_{lp}\sum_{\sigma} \dop{c}{l p\sigma} \op{d}_{\sigma} + \mbox{h.c.} \,,\nn
\end{align}
where $\mathcal{H}_{\mbs{J}}$ and $\mathcal{H}_{\mbs{int}}$ have been defined
in (\ref{eq:hamiltonian}).
Using the EOM technique introduced in Sec.~\ref{sec:semiclass} and the
Lindblad master equation for 
the SQD setup
\begin{align}
\frac{\d}{\d t}\op{\rho}(t) &= -\imath \comm{\mathcal{H}_{\mbs{SQD}}}{\op{\rho}(t)}\nnn 
&- \frac{1}{2}  \sum_\sigma \Bigg[ \Gamma_{L\sigma} \left(\anticomm{\op{d}_{\sigma} \dop{d}{\sigma}}{\op{\rho}(t)} - 2 \dop{d}{\sigma} \op{\rho}(t) \op{d}_{\sigma} \right) \nnn 
& \quad \quad + \Gamma_{R\sigma} \left(\anticomm{\dop{d}{\sigma} \op{d}_{\sigma}}{\op{\rho}(t)} - 2 \op{d}_{\sigma} \op{\rho}(t) \dop{d}{\sigma} \right) \Bigg]\,,
\end{align}
we obtain the following
EOM  for the QD occupations, the electron spin, and the large spin 
($\Gamma \equiv \Gamma_L = \Gamma_R$, omitting the time dependence):

\begin{align}
 \frac{\d}{\d t} \expval{\op{n}^\sigma} &= \left(\lambda^x\expval{\op{J}^x} \expval{\op{S}^y}-\lambda^y\expval{\op{J}^y} \expval{\op{S}^x}\right) \left(\delta_{\sigma\spup}-\delta_{\sigma\spdo}\right) \nnn 
 &\quad-(\Gamma_{R\sigma}+\Gamma_{L\sigma}) \expval{\op{n}^\sigma} + \Gamma_{L\sigma} \nnn
\frac{d}{dt}\expval{\op{\vec{S}}}&=\big[\expval{\op{\vec{J}}'}+B\vec{e}_z\big]\times\expval{\op{\vec{S}}}
-\Gamma\expval{\op{\vec{S}}} \nnn 
& \quad - \frac{1}{2}\Gamma(p_L+p_R)\expval{\op{N}} \,\vec{e}_z +\frac{1}{2}\Gamma\, p_L\,
\vec{e}_z\,,\nnn
\frac{d}{dt}\expval{\op{\vec{J}}}&=\big[\expval{\op{\vec{S}}'}+B\vec{e}_z\big]\times\expval{\op{\vec{J}}}
\label{eq:single_eom}
\end{align}
with $\expval{\vec{\op{S}}'} \equiv \sum_\alpha\lambda^\alpha \expval{\op{S}^{\alpha}}\vec{e}_\alpha$
, $\expval{\vec{\op{J}}'} \equiv\sum_\alpha\lambda^\alpha \expval{\op{J}^{\alpha}}\vec{e}_\alpha$ and $\expval{\op{N}} = \expval{\op{n}_\spup}+\expval{\op{n}_\spdo}$.
The anisotropic version of these equations ($\lambda^x = \lambda^z = \lambda$  and
$\lambda^y = 0$) has been already studied in
Ref.~\cite{Lopez-Monis2012}, but with a different focus.
For isotropic coupling Fig.~\ref{fig:pleft_pright_currents_switching_contour}(b) presents the
steady-state current and the final
direction of the large spin depicted as vertical arrows in dependence on the
lead polarizations $p_L$ and $p_R$. 
The transition between the two directions occurs for $p_L=p_R$,
which has been also obtained in  Ref.~ \cite{Bode2012}.
There the magnetization change  was discussed in terms of the sign of the spin-transfer
torque, which is determined by 
$\textrm{sgn}[\Gamma_{L\spdo} \Gamma_{R\spup} - \Gamma_{L\spup}
\Gamma_{R\spdo}]$.
In contrast to the switching in DQD 
[see Fig.~\ref{fig:pleft_pright_currents_switching_contour}(a)] the
transition does not depend on $\lambda$ or $J$.
We further note that since the electrons are assumed to be noninteracting
in the single-QD the current is symmetric with respect to an exchange
of $p_L$ and $p_R$.
This does not hold for the DQD.

The comparison of the current evolutions for both setups (as shown in
Fig.~\ref{fig:compare_switching_SQD_DQD_currents}) reveals that
the SQD switching occurs always faster for the same set of parameters.
In other words the coherent electron transfer between the QDs slows down the switching.

\begin{figure}[htb]
  \centering
  \includegraphics[width=0.5\textwidth,keepaspectratio=True]{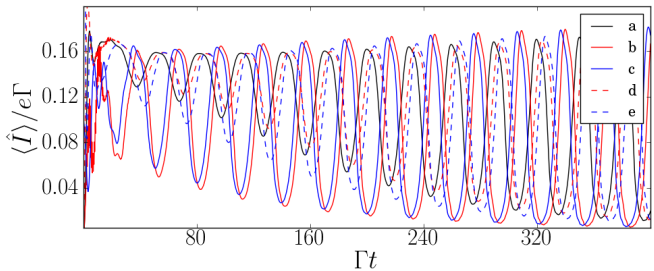}
  \caption{(Color online) Time-dependent currents in the case of undamped large spin for different initial system state: (a) empty DQD , (b) $n_{L \spup}(0) = 1$ , (c) $n_{L \spdo}(0) = 1$, (d) $n_{R \spup}(0) = 1$ and (e) $n_{R \spdo}(0) = 1$. The dominating oscillation frequencies are multiples of $2B$ for all currents (parametric resonance; see text),
but we observe phase shifts and different magnitudes.
Parameters: $p_L = p_R = -0.9$, $B/\Gamma =$~0.1, $T_c/\Gamma =$~0.4, $\lambda_L^\alpha = \lambda_R^x = \lambda_R^z = \Gamma$, $\lambda_R^y = 0$. }
  \label{fig:currents_anisotrop_undamped_diff_init} 
 \end{figure}

\subsection{\label{sec:anisotrop}Anisotropic coupling}

{\it Large spin switching.} --
Given the investigations of the spin magnetization reversal in the isotropic case we probe for anisotropic coupling schemes that yield spin switching as well.  
In principle we observe that the switching process takes place as long as the $x$-and $y$-components of the spins are isotropically 
coupled ($\lambda_i^x \approx \lambda_i^y$, $\forall i$) and the leads polarized suitably. 
Taking into account the microscopic spin interaction [see Sec.~\ref{sec:isotrop}]  this becomes clear since the $\lambda^z_i$ terms are not directly contributing in the 
spin-transfer process.
However, the $z-$couplings affect the speed of magnetization reversal:
The magnetization process, e.g., can be slowed down by increasing the $z-$ couplings on both sites: $\lambda_{L}^z = \lambda_{R}^z \equiv \lambda^z > \lambda$.
Contrariwise, decreasing $\lambda^{z}$ steps up the switching.
Even if the $z$-couplings are different switching occurs. 
The corresponding currents display a peak at the time when the derivative of $\expval{\op{J}^z}(t)$ is maximal. 
%

\begin{figure*}[t!]
\centering
\subfloat{
\includegraphics[width=0.48\textwidth,keepaspectratio=true]{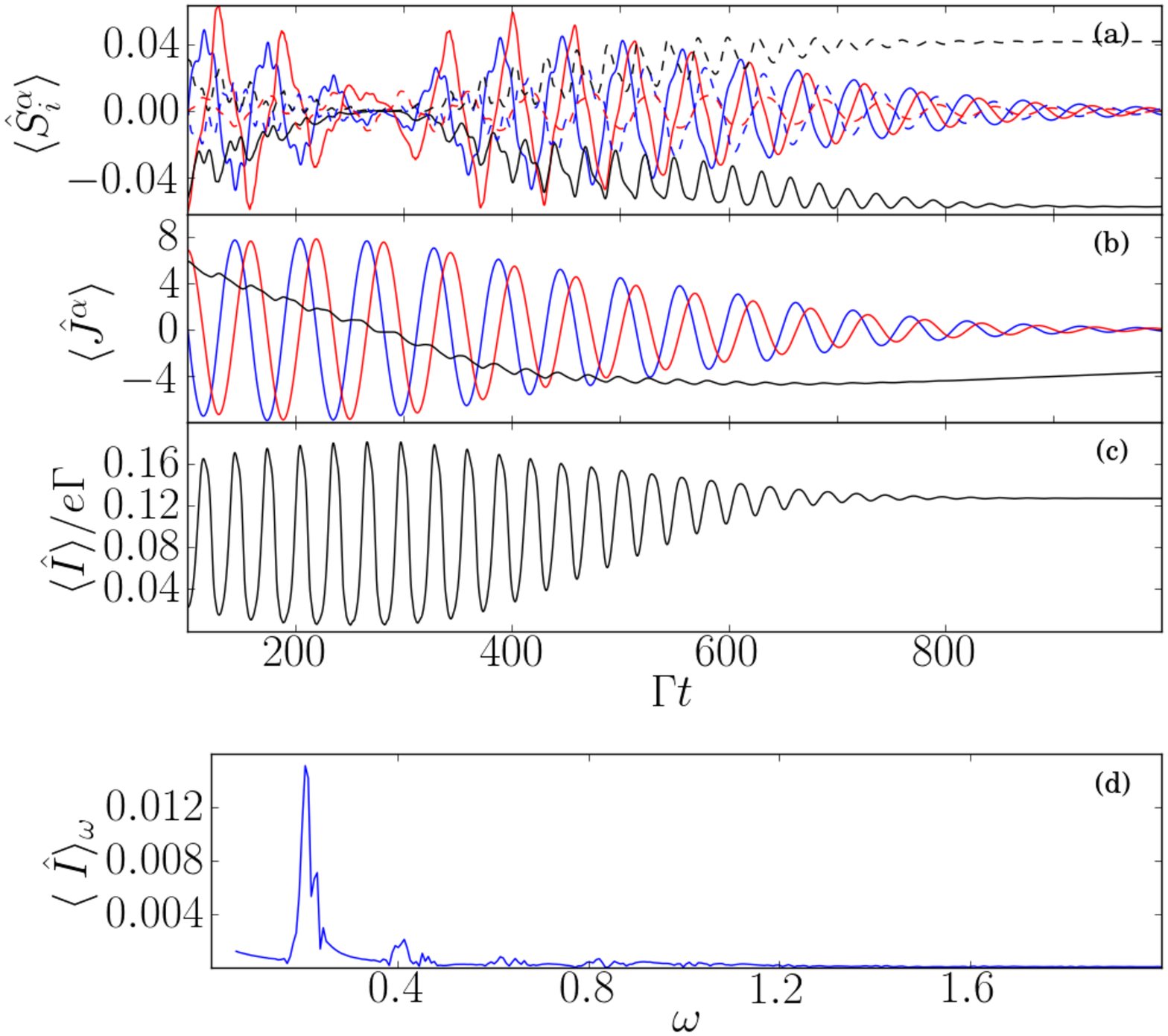}
}
\subfloat{
\includegraphics[width=0.48\textwidth,keepaspectratio=true]{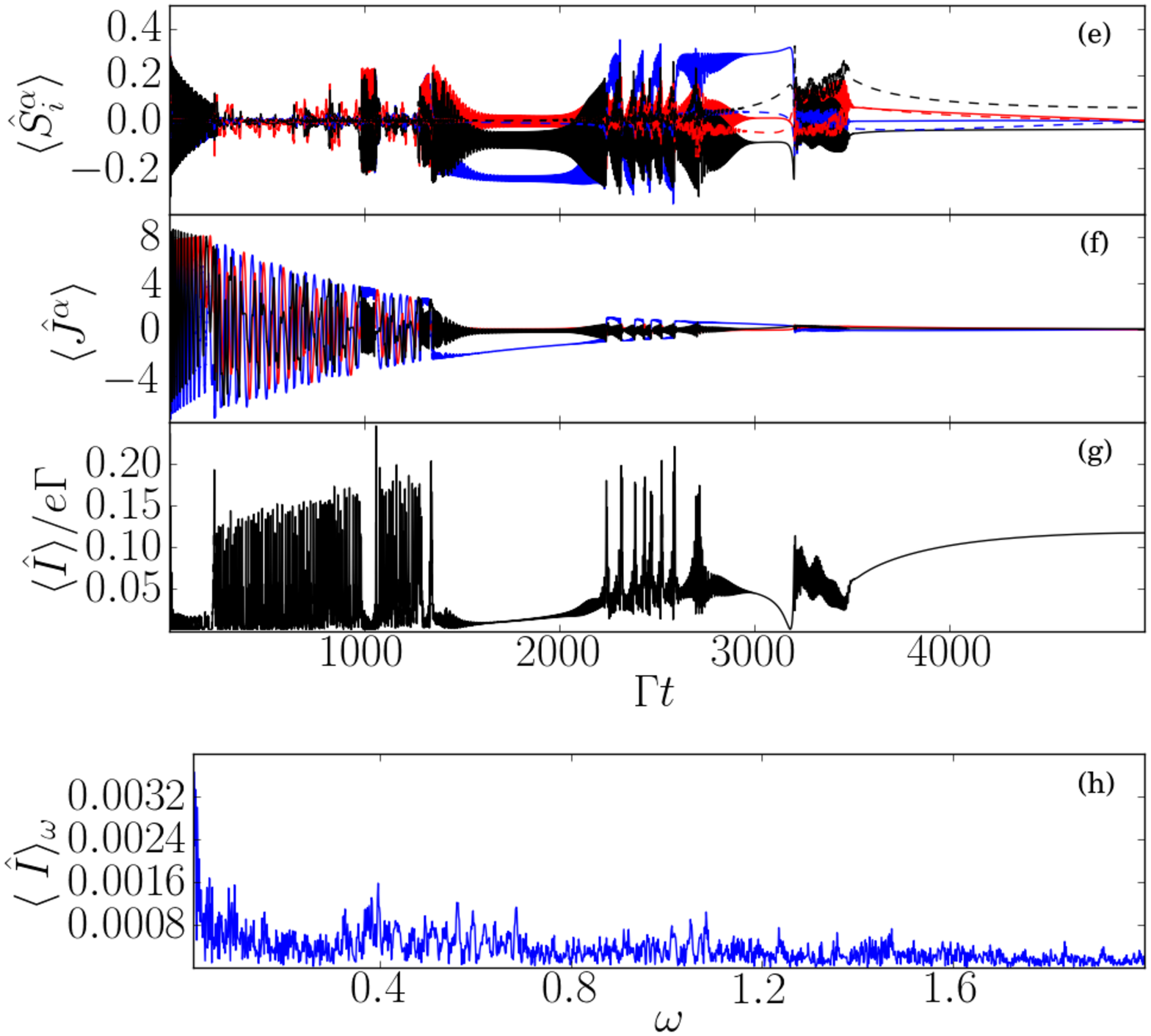}
}
 \caption{(Color online) Different transient currents for a setup with strongly polarized contacts ($\,p_L = p_R = -0.9$ and $T_c/\Gamma = 0.4$, $B/\Gamma = 0.1$, $\varepsilon = 0$,
$\gamma_J/\Gamma  = 10^{-3} $) and different couplings between spins: (Left) $\lambda_L^\alpha = \lambda_R^x = \lambda_R^z = \Gamma$, $\lambda_R^y = 0$ and (Right)
$\lambda_L^x/\Gamma = 0.2,\,\lambda_L^y/\Gamma = 4.2,\,\lambda_L^z/\Gamma = 1.3,\,\lambda_R^x/\Gamma = 2.9\,,\lambda_R^y/\Gamma = 1.5,\,\lambda_R^z/\Gamma = 0.7$.
(a) and (e) Components of the electron spin $\expval{\op{S}_{L}^x}$ (blue, solid), $\expval{\op{S}_{R}^x}$ (blue, dashed)
$\expval{\op{S}_{L}^y}$ (red, solid), $\expval{\op{S}_{L}^y}$ (red, dashed), $\expval{\op{S}_{L}^z}$ (black, solid), $\expval{\op{S}_{R}^z}$ (black, dashed). 
(b) and (f) Components of the large spin: $\expval{\op{J}^x}$ (blue),  $\expval{\op{J}^y}$ (red),  $\expval{\op{J}^z}$ (black).
(c) and (g) Electronic current $\expval{I}(t)$ vs. time.
The steady-state current is given by Eq.~\eqref{eq:stationary_current}.
(d) and (h) In the current spectrum of (c) the dominating frequencies are even multiples (2,4,6\dots) of the large spin Larmor frequency $B$,
which refers to the phenomenon of parametric resonance (see text).
The current spectrum of (g) is broad and noisy, which we attribute to chaotic behavior (see also Ref.~\cite{Lopez-Monis2012}).
For both cases holds: As the large spin is damped out, the current oscillations vanish, since no more spin-flips occur.}
\label{fig:transients_anisotropy}
\end{figure*} 

\noindent
{\it Non-magnetic leads $p_L=p_R=$~0. }-- Basically the same as for the isotropic coupling holds. 
In addition we find that the transient phase and, thus, the polar angle for the free precession depend crucially on the initial electronic states.

{\it Undamped large spin $\gamma_J=$~0.} -- We obtain various regimes with respect to lead polarization and spin coupling that exhibit limit cycles in the current, similar to
Ref.~\cite{Lopez-Monis2012}.
The Fourier spectra of those current time evolutions exhibit peaks at well defined frequencies, confirming the periodicity of the system's evolution. 
We note that in contrast to the single-QD\cite{Lopez-Monis2012} the initial system occupation plays an important r\^ole in the evolution of both the DQD and the
large spin, which is illustrated in Fig.~\ref{fig:currents_anisotrop_undamped_diff_init}.
A similar observation  has been made in Ref.~\cite{ERL05}.
For a more detailed analysis the fix points of the EOM have to be determined.
Unfortunately, due to the size and complexity of the EOM, we did not manage to find them analytically.
Even their numerical computation  provides a complicated task. 

{\it Parametric resonance.} --
Studying the rich dynamics provided by the anisotropy, we can observe
parametric resonance in the current oscillations, particularly. 
The evolutions exhibit current oscillations at multiples of the
doubled Larmor frequency of the 
large spin, i.e., $2 B$, as can be seen in Figs.~\ref{fig:transients_anisotropy}(a)--~\ref{fig:transients_anisotropy}(c).
Although the periodic oscillations are in general very complicated,
due to the nonlinearity in
the equations, the dominating frequency of the oscillations of the
$x$ and $y$ components of the
electron spins, respectively, is $B$. 
The $z$ component and consequently the current, on the other hand,
oscillates with multiples of $2 B$, which we reveal in the
Fourier spectrum [Fig.~\ref{fig:transients_anisotropy}(d)].
The same phenomenon has been observed in the single-QD, so that
  we can employ its model to provide an explanation.
We assume anisotropic 
spin-spin coupling, i.e., $\lambda^x = \lambda^z = \lambda$  and
$\lambda^y = 0$
and magnetic leads ($p_L,p_R\neq$~0).
Given that $\lambda J\ll$~1 we can make use of  the
back-action free EOM for the large spin \eqref{eq:precess}.
It follows from Eqs.~\eqref{eq:single_eom} that $\expval{\op{S}^x}$
and $\expval{\op{S}^y}$ oscillate with frequency $B$.
We further recognize that the time derivative of 
$\expval{\op{n}^\sigma}$ couples to the 
product $\expval{\op{S}^y} \expval{\op{J}^x}$.
Integrating this product of sinusoidal oscillations both with frequency $B$
leads us to the current and $\expval{\op{S}^z}$ oscillating with
frequency $2B$ \cite{Metelmann2012}.

Given that the large spin is damped by the rate $\gamma_J$ the evolutions
for different 
initial DQD occupations match in the long-time limit. 
Consequently, the periodic or nonperiodic oscillations are only
present during the transients (as shown in the panels of Fig.~\ref{fig:transients_anisotropy}).
We observe different transient scenarios ranging from quasiperiodic
[(a)--(d)] to chaotic motion [(e)--(h)].
In all cases the steady-state current is given by \eqref{eq:stationary_current}. 

\section{\label{sec:conclude}Conclusions}

In this work we have studied electron transport through a DQD setup
coupled to electronic contacts . 
The spin of the excess electrons in the DQD interacts with a large
spin and an external 
magnetic field is applied.
We use semiclassical Ehrenfest EOM together with a quantum master
equation technique.
This method works well if one assumes that the dwell time of the
electrons is much smaller than the average large spin precession period.

We have found that the coupled dynamics of the large spin and the electron spins 
as well as 
the current through the DQD structure strongly depend on the
polarization 
of the electronic leads and on the isotropy of the spin-spin coupling.
Particularly, if the coupling between electron spin and large spin is 
isotropic with respect to $x$ and $y$  while the electronic leads are reversely polarized,
a current-induced magnetization process of 
the large spin is obtained. 
The speed of this switching is always lower than for a single-QD
setup and depends, most importantly, on the lead polarization.

On the other hand, a complete anisotropic coupling leads to a rich 
variety of different dynamical scenarios, which for an undamped large
spin depend on the initial state.
In particular, we have found that the electronic system may show
self-sustained oscillations, 
either quasi-periodic or even chaotic. 
Introducing a large spin damping renders these dynamics to be
transient.
A remarkable feature of the undamped or damped dynamics is the occurence of parametric
resonance, observable as frequency doubling in the current. 
Furthermore, we analytically have derived the stationary currents for the
spin-dependent 
unidirectional single-electron transport, 
when no interactions with the external spin are present. 
It turns out that this current also applies for the case of current-polarized or damped large spin.

Experimental realizations for the large spin could be either the spin of
magnetic impurities in semiconductor QDs or a net magnetic moment in
single molecules (single molecular magnets). 
Whether our mean-field approach is applicable for these examples, where
the spin typically is not very large, needs further investigations.
However, one possible realization for the large spin, that might justify the
semi-classical treatment is the hyperfine interaction with an ensemble
of nuclear spins, where the number of spins is reasonably high.\cite{Schuetz2012}


\begin{acknowledgments}
The authors would like to thank the DFG (BR 1528/8 and SFB 910) for financial support.
Discussions with Carlos L\'opez-Mon\'is and Anja Metelmann are gratefully acknowledged.
\end{acknowledgments}



\appendix
\renewcommand{\theequation}{\thesection.\arabic{equation}}
\section{\label{sec:block_matrices}Block-matrices and parameters}

Here, we provide the coupling matrices used in the EOM for the electronic part (\ref{eq:mastereq})
\begin{align}
 \mat{L}^\sigma \left(\expval{\op{J}^z} \right) &= 
 \begin{pmatrix}
  0 & 0 & -\imath T_c & \imath T_c \\
  0 & -\Gamma_{R\sigma} & \imath T_c & -\imath T_c \\
  -\imath T_c & \imath T_c & -\frac{\Gamma_{R\sigma}}{2} + \imath \varepsilon^\sigma & 0 \\
  \imath T_c & -\imath T_c & 0 & -\frac{\Gamma_{R\sigma}}{2} - \imath \varepsilon^\sigma
 \end{pmatrix} \nonumber
\end{align}
\begin{align}
 \mat{A} \left(\expval{\op{J}^x},\expval{\op{J}^y} \right) &= \begin{pmatrix}
	    -\Lambda_L^- & \Lambda_L^+ & 0 & 0 \\
	    0 & 0 & -\Lambda_R^- & \Lambda_R^+ \\
	    0 & 0 & 0 & 0 \\
	    0 & 0 & 0 & 0
           \end{pmatrix}\nonumber
\end{align}
\begin{align}
 \mat{B} \left(\expval{\op{J}^x},\expval{\op{J}^y} \right) = \begin{pmatrix}
	    0 & 0 & 0 & 0 \\
            0 & 0 & 0 & 0 \\
	    -\Lambda_R^- & 0 & 0 & \Lambda_L^+ \\
	    0 & \Lambda_R^+  & -\Lambda_L^- & 0 
           \end{pmatrix} \nonumber
\end{align}
\begin{align}
\mat{C} \left(\expval{\op{J}^x},\expval{\op{J}^y} \right) &= \begin{pmatrix}
	    0 & 0 & 0 & 0 \\
            0 & 0 & 0 & 0 \\
	    \Lambda_L^- & 0 & 0 & -\Lambda_R^+ \\
	    0 & -\Lambda_L^+  & \Lambda_R^- & 0 
           \end{pmatrix}\nonumber
\end{align}
\begin{align}
 \mat{E} = \begin{pmatrix}
        -\imath T_c & 0 & \imath T_c & 0 \\
	0 & \imath T_c & 0 & -\imath T_c \\
        \imath T_c & 0 & -\imath T_c & 0 \\
	0 & -\imath T_c & 0 & \imath T_c
       \end{pmatrix}\nonumber
\end{align}
\begin{align}
 \mat{D} \left(\expval{\op{J}^z} \right) &= \textrm{Diag}\bigg(
        \imath \tilde{\varepsilon}_L, -\imath \tilde{\varepsilon}_L, \imath \tilde{\varepsilon}_R -\frac{\Gamma_{R}}{2},-\imath \tilde{\varepsilon}_R -\frac{\Gamma_{R}}{2}
       \bigg)\nnn
 \mat{F} \left(\expval{\op{J}^z} \right) &= \textrm{Diag}\bigg(
        \imath \varepsilon'_+ -\frac{\Gamma_{R\spdo}}{2}, -\imath \varepsilon'_+ -\frac{\Gamma_{R\spdo}}{2}, \imath \varepsilon'_- -\frac{\Gamma_{R\spup}}{2},
\nnn 
& -\imath \varepsilon'_- -\frac{\Gamma_{R\spup}}{2}
       \bigg)\nonumber
\end{align}
with the parameters:
\begin{align}
\label{eq:matrix_parameters}
\varepsilon^\sigma &\equiv \varepsilon + \frac{1}{2} (\delta_{\sigma \spup} - \delta_{\sigma \spdo}) \left(\lambda_L^z-\lambda_R^z\right) \expval{\op{J}^z} \,,\nnn
\varepsilon'_{\pm} &\equiv \pm \varepsilon + B + \frac{1}{2} \left(\lambda_L^z-\lambda_R^z\right) \expval{\op{J}^z} \,,\\
\tilde{\varepsilon}_i &\equiv B + \lambda_i^z \expval{\op{J}^z} \,, (i=L,R) \nnn
\Lambda_i^{\pm} &\equiv \frac{\imath}{2} \left[\lambda_i^x \expval{\op{J}^x} \pm \imath \lambda_i^y \expval{\op{J}^y}\right] \quad ; 
\quad \left(\Lambda_i^{\pm}\right)^* = - \Lambda_i^{\mp} \,. \nn
\end{align}


\section{\label{sec:no_backaction}Large-spin dynamics with vanishing electronic back-action}

Without the electronic spin the EOM  of the large spin \eqref{eq:eom_largespin} or in \eqref{eq:single_eom} read 
\begin{align}
\frac{\d}{\d t} \expval{\vec{\op{J}}} &= \vec{B} \times \expval{\vec{\op{J}}} - \gamma_J \expval{\vec{\op{J}}}\,,
\end{align}
which are readily solved by
\begin{align}
\vec{\expval{\op{J}}}(t) &= \e^{-\gamma_J t}
    \begin{pmatrix} \cos (B t) & - \sin (B t) & 0 \\
      \sin (B t) & \cos (B t) & 0 \\
      0 & 0 & 1
    \end{pmatrix} \expval{\vec{\op{J}}}_0\,.
\label{eq:precess}
\end{align}
This provides the damped Larmor precession of the large spin around the external magnetic field axis with frequency $B$.
Inserting Eqs.~\eqref{eq:precess} into the electronic EOMs leads to a set of linear first order differential equations with time-periodic
coefficients.
In two dimensions it corresponds to the Mathieu-Hill equation, which describes the phenomenon of parametric resonance.



\end{document}